\documentclass[10pt,twocolumn]{article}
\usepackage{amssymb}
\usepackage{amsmath}

\input{tcilatex}

\begin{document}

\title{Electromagnetic Field in Lyra Manifold: A First Order Approach}
\author{R. Casana, C. A. M. de Melo and B. M. Pimentel \\
{\small Instituto de F\'{\i}sica Te\'orica, Universidade Estadual Paulista} 
\vspace{-.1cm}\\
{\small Rua Pamplona 145, CEP 01405-900, S\~{a}o Paulo, SP, Brazil}\\
{\small \textbf{Abstract:} We discuss the coupling of the electromagnetic
field with a curved}\\
{\small and torsioned Lyra manifold using the Duffin-Kemmer-Petiau theory.
We will show}\\
{\small how to obtain the equations of motion and energy-momentum and spin
density tensors}\\
{\small by means of the Schwinger Variational Principle.}}
\date{}
\maketitle


\section{Introduction}

First order Lagrangians are one of the most profitable tools in Field
Theory. By means of first order approach, Hamiltonian dynamics becomes more
transparent, constrained systems can be dealt with a wide range of methods 
\cite{FirstOrderConstrained}, and CPT and spin-statistics theorems can be
proved by variational statements \cite{SchwSpinStat}.

Otherwise, the coupling between electromagnetism and the torsion content of
spacetime has been \ an intringuing puzzle for many years. Minimal coupling
of the Einstein-Cartan gravity with eletromagnetism breaks local gauge
covariance by the presence of the torsion interaction \cite{sabbata,
RiemCart, MassiveRCartan}.

Here, we want to add another piece to the puzzle, showing that the torsion
coupling problem is related to scale invariance which we will model together
with the gravitational field by means of the Lyra geometry. Electromagnetic
field will be described by the first order approach of Duffin-Kemmer-Petiau
(DKP).

\section{The Lyra Geometry\label{LyraGeom}}


The Lyra manifold \cite{Lyra} is defined giving a tensor metric $g_{\mu \nu
} $ and a positive definite scalar function $\phi $ which we call the scale
function. In Lyra geometry, one can change scale and coordinate system in an
independent way, to compose what is called a \emph{reference system}
transformation: let $M\subseteq \mathbb{R}^{N}$ and $U$ an open ball in $%
\mathbb{R}^{n}$, ($N\geq n$) and let $\chi :U\curvearrowright M$. The pair $%
\left( \chi ,U\right) $ defines a \emph{coordinate system}. Now, we define a
reference system by $\left( \chi ,U,\phi \right) $ where $\phi $ transforms
like 
\begin{equation*}
\bar{\phi}\left( \bar{x}\right) =\bar{\phi}\left( x\left( \bar{x}\right)
;\phi \left( x\left( \bar{x}\right) \right) \right) \quad ,\quad \frac{%
\partial \bar{\phi}}{\partial \phi }\not=0
\end{equation*}%
under a reference system transformation.

In the Lyra's manifold, vectors transform as 
\begin{equation*}
\bar{A}^{\nu }=\frac{\bar{\phi}}{\phi }\frac{\partial \bar{x}^{\nu }}{%
\partial x^{\mu }}A^{\mu }
\end{equation*}%
In this geometry, the affine connection is 
\begin{equation*}
\tilde{\Gamma}_{\;\,\mu \nu }^{\rho }=\frac{1}{\phi }\mathring{\Gamma}^{\rho
}{}_{\mu \nu }+\frac{1}{\phi }\left[ \delta _{\,\mu }^{\rho }\partial _{\nu
}\ln \left( \frac{\phi }{\bar{\phi}}\right) -g_{\mu \nu }g^{\rho \sigma
}\partial _{\sigma }\ln \left( \frac{\phi }{\bar{\phi}}\right) \right]
\end{equation*}%
whose transformation law is given by 
\begin{eqnarray*}
\tilde{\Gamma}_{\;\,\mu \nu }^{\rho } &=&\frac{\bar{\phi}}{\phi }\bar{\Gamma}%
_{\;\,\lambda \varepsilon }^{\sigma }\frac{\partial x^{\rho }}{\partial \bar{%
x}^{\sigma }}\frac{\partial \bar{x}^{\lambda }}{\partial x^{\mu }}\frac{%
\partial \bar{x}^{\varepsilon }}{\partial x^{\nu }}+\frac{1}{\phi }\frac{%
\partial x^{\rho }}{\partial \bar{x}^{\sigma }}\frac{\partial ^{2}\bar{x}%
^{\sigma }}{\partial x^{\mu }\partial x^{\nu }}+ \\
&&+\frac{1}{\phi }\delta _{\,\nu }^{\rho }\frac{\partial }{\partial x^{\mu }}%
\ln \left( \frac{\bar{\phi}}{\phi }\right) \,.
\end{eqnarray*}

One can define the covariant derivative for a vector field as 
\begin{equation*}
\nabla _{\mu }A^{\nu }={\frac{1}{\phi }}\partial _{\mu }A^{\nu }+\tilde{%
\Gamma}_{\;\,\mu \alpha }^{\nu }A^{\alpha }\,,\quad \nabla _{\mu }A_{\nu }={%
\frac{1}{\phi }}\,\partial _{\mu }A_{\nu }-\tilde{\Gamma}_{\;\,\mu \nu
}^{\alpha }A_{\alpha }\,.
\end{equation*}%
We use the notation $\Gamma _{\;\,[\alpha \mu ]}^{\nu }=\frac{1}{2}\left(
\Gamma _{\;\alpha \mu }^{\nu }-\Gamma _{\;\mu \alpha }^{\nu }\right) $ for
the antisymmetric part of the connection and $\mathring{\Gamma}_{\,\mu \nu
}^{\rho }\equiv \frac{1}{2}g^{\rho \sigma }\left( \partial _{\mu }g_{\nu
\sigma }+\partial _{\nu }g_{\sigma \mu }-\partial _{\sigma }g_{\mu \nu
}\right) $ for the analogous of the Levi-Civita connection.

The richness of the Lyra's geometry is demonstrated by\ the \emph{curvature} 
\cite{Sen} 
\begin{eqnarray*}
\tilde{R}_{\,\beta \alpha \sigma }^{\rho } &\equiv &\frac{1}{\phi ^{2}}\left[
\partial _{\beta }\left( \phi \tilde{\Gamma}_{\,\alpha \sigma }^{\rho
}\right) -\partial _{\alpha }\left( \phi \tilde{\Gamma}_{\,\beta \sigma
}^{\rho }\right) \right] + \\
&&+\frac{1}{\phi ^{2}}\left[ \phi \tilde{\Gamma}_{\,\beta \lambda }^{\rho
}\phi \tilde{\Gamma}_{\,\alpha \sigma }^{\lambda }-\phi \tilde{\Gamma}%
_{\,\alpha \lambda }^{\rho }\phi \tilde{\Gamma}_{\,\beta \sigma }^{\lambda }%
\right]
\end{eqnarray*}%
and the \emph{torsion} 
\begin{equation}
\tilde{\tau}_{\mu \nu }^{\,\quad \rho }=-\frac{2}{\phi }\delta _{\,[\mu
}^{\rho }\partial _{\nu ]}\ln \bar{\phi}  \label{torsion1}
\end{equation}%
which has intrinsic link with the scale functions and whose trace is given
by 
\begin{equation}
\tilde{\tau}_{\mu \rho }^{\,\quad \rho }\equiv \tilde{\tau}_{\mu }=\ \frac{3%
}{\phi }\partial _{\mu }\ln \bar{\phi}\,.  \label{torsion-trace}
\end{equation}

In the next section we introduce the behavior of massless DKP field in the
Lyra geometry.

\section{The massless DKP field in Lyra manifold}

DKP theory describes in a unified way the spin $0$ and $1$ fields \cite%
{DKPoriginais, KrajcikNieto, Lunardi1}. The massless DKP theory can not be
obtained as a zero mass limit of the massive DKP case, so we consider the
Harish-Chandra Lagrangian density for the massless DKP theory in the
Minkowski space-time $\mathcal{M}^{4}$, given by \cite{HarishC} 
\begin{equation}
\mathcal{L_{M}}=i\bar{\psi}\gamma \beta ^{a}\partial _{a}\psi -i\partial _{a}%
\bar{\psi}\beta ^{a}\gamma \psi -\bar{\psi}\gamma \psi \;,  \label{minkowski}
\end{equation}%
where the $\beta ^{a}$ matrices satisfy the usual DKP algebra 
\begin{equation*}
\beta ^{a}\beta ^{b}\beta ^{c}+\beta ^{c}\beta ^{b}\beta ^{a}=\beta ^{a}\eta
^{bc}+\beta ^{c}\eta ^{ba}
\end{equation*}%
and $\gamma $ is a \textit{singular} matrix satisfying\footnote{%
We choose a representation in which ${\beta ^{0}}^{\dag }={\beta ^{0}}$, ${%
\beta ^{i}}^{\dag }=-{\beta ^{i}}$ and $\gamma ^{\dag }=\gamma $ \thinspace .%
} 
\begin{equation*}
\beta ^{a}\gamma +\gamma \beta ^{a}=\beta ^{a}\qquad ,\qquad \gamma
^{2}=\gamma \,.
\end{equation*}

From the above lagrangian it follows the massless DKP wave equation 
\begin{equation*}
i\beta ^{a}\partial _{a}\psi -\gamma \psi =0\;.
\end{equation*}

As it was known, the Minkowskian Lagrangian density (\ref{minkowski}) in its
massless spin 1 sector reproduces the electromagnetic or Maxwell theory with
its respective $U(1)$ local gauge symmetry.

To construct the covariant derivative of massless DKP field in Lyra
geometry, we follow the standard procedure of analyzing the behavior of the
field under local Lorentz transformations, 
\begin{equation}
\psi \left( x\right) \rightarrow \psi \prime \left( x\right) =U\left(
x\right) \psi \left( x\right)  \label{LocalLorentz}
\end{equation}%
where $U$\ is a spin representation of Lorentz group characterizing the DKP\
field. Now we define a \emph{spin connection} $S_{\mu }$ in a such way that
the object 
\begin{equation}
\nabla _{\mu }\psi \equiv \frac{1}{\phi }\partial _{\mu }\psi +S_{\mu }{}\psi
\label{cov-fer}
\end{equation}%
transforms like a DKP field in (\ref{LocalLorentz}), thus, we set%
\begin{equation*}
\nabla _{\mu }\psi \rightarrow \left( \nabla _{\mu }\psi \right) ^{\prime
}=U\left( x\right) \nabla _{\mu }\psi
\end{equation*}%
and therefore $S$ transforms like 
\begin{equation}
S_{\mu }^{\prime }=U\left( x\right) S_{\mu }U^{-1}\left( x\right) -\frac{1}{%
\phi }\left( \partial _{\mu }U\right) U^{-1}\left( x\right)
\label{TransfSpin}
\end{equation}

From the covariant derivative of the DKP field (\ref{cov-fer}) and
remembering that $\bar{\psi}\psi $ must be a scalar under the transformation
(\ref{LocalLorentz}), it follows that $\nabla _{\mu }\bar{\psi}=\frac{1}{%
\phi }\partial _{\mu }\bar{\psi}-\bar{\psi}S_{\mu }$. Using the covariant
derivative of the DKP current 
\begin{gather*}
\nabla _{\mu }\left( \bar{\psi}\beta ^{\nu }\psi \right) =\frac{1}{\phi }%
\partial _{\mu }\left( \bar{\psi}\beta ^{\nu }\psi \right) +\Gamma ^{\nu
}{}_{\mu \lambda }\left( \bar{\psi}\beta ^{\lambda }\psi \right) = \\
=\left( \nabla _{\mu }\bar{\psi}\right) \beta ^{\nu }\psi +\bar{\psi}\left(
\nabla _{\mu }\beta ^{\nu }\right) \psi +\bar{\psi}\beta ^{\nu }\left(
\nabla _{\mu }\psi \right)
\end{gather*}%
one gets the following expression for the covariant derivative of $\beta
^{\nu }$%
\begin{equation*}
\nabla _{\mu }\beta ^{\nu }=\frac{1}{\phi }\partial _{\mu }\beta ^{\nu
}+\Gamma ^{\nu }{}_{\mu \lambda }\beta ^{\lambda }+S_{\mu }\beta ^{\nu
}-\beta ^{\nu }S_{\mu }
\end{equation*}

A particular solution to this equation is given by 
\begin{equation*}
S_{\mu }=\frac{1}{2}\omega _{\mu ab}S^{ab}~,\;S^{ab}=\left[ \beta ^{a},\beta
^{b}\right] .
\end{equation*}

With a covariant derivative of the DKP\ field well-defined we can consider
the Lagrangian density (\ref{minkowski}) of the massless DKP field minimally
coupled \cite{misner, hehl, sabbata} to the Lyra manifold, introducing the
tetrad field,%
\begin{equation*}
g^{\mu \nu }(x)=\eta ^{ab}\,e^{\mu }{}_{a}(x)e^{\nu }{}_{b}(x)\,,\;g_{\mu
\nu }(x)=\eta _{ab}e_{\mu }{}^{a}(x)e_{\nu }{}^{b}(x)~.
\end{equation*}%
\begin{equation}
S=\int_{\Omega }\!\!d^{4}x\;\phi ^{4}e\;\left( \!i\,\bar{\psi}\gamma
e_{\;\,a}^{\mu }\beta ^{a}\nabla _{\mu }\psi -i\nabla _{\mu }\bar{\psi}\beta
^{a}e_{\;\,a}^{\mu }\gamma \psi -\bar{\psi}\gamma \psi \right) \;.
\label{lag-cv}
\end{equation}%
where $\nabla _{\mu }$ is the Lyra covariant derivative of DKP field defined
above.

\section{Equations of Motion and the Description of Matter Content}

In following we use a classical version of the Schwinger Action Principle
such as it was treated in the context of Classical Mechanics by Sudarshan
and Mukunda \cite{Sudarshan}. The Schwinger Action Principle is the most
general version of the usual variational principles. It was proposed
originally at the scope of the Quantum Field Theory \cite{SchwSpinStat}, but
its application goes beyond this area. Here, we will apply the Action
Principle to derive equations of motion of the Dirac field in an external
Lyra background and expression for the energy-momentum and spin density
tensors.

Thus, making the variation of the action integral (\ref{lag-cv}) we get%
\begin{equation*}
\delta S=\int_{\Omega }dx~e\phi ^{4}\left[ 4\mathcal{L}-\frac{i}{\phi }\bar{%
\psi}\gamma \beta ^{\mu }\partial _{\mu }\psi +\frac{i}{\phi }\partial _{\mu
}\bar{\psi}\beta ^{\mu }\gamma \psi \right] \left( \frac{\delta \phi }{\phi }%
\right) +
\end{equation*}%
\begin{gather}
+\int_{\Omega }dx~\phi ^{4}e~\left( \frac{\delta e}{e}\right) \mathcal{L~}+
\label{action-variation} \\
+\int_{\Omega }dx~e\phi ^{4}\left[ \frac{{}}{{}}i\bar{\psi}\gamma \left(
\delta \beta ^{\mu }\right) \nabla _{\mu }\psi -i\nabla _{\mu }\bar{\psi}%
\left( \delta \beta ^{\mu }\right) \gamma \psi \right] +  \notag \\
+\int_{\Omega }dx~e\phi ^{4}\left[ \frac{{}}{{}}i\bar{\psi}\gamma \beta
^{\mu }\left( \delta S_{\mu }\right) \psi +i\bar{\psi}\left( \delta S_{\mu
}\right) \beta ^{\mu }\gamma \psi \right] +  \notag \\
+\int_{\Omega }dx~e\phi ^{4}~\delta \bar{\psi}\left( i\gamma \beta ^{\mu
}\nabla _{\mu }\psi -\gamma \psi +iS_{\mu }\beta ^{\mu }\gamma \psi \right) +
\notag \\
+\int_{\Omega }dx~e\phi ^{4}\left[ \frac{i}{\phi }\bar{\psi}\gamma \beta
^{\mu }\left( \delta \partial _{\mu }\psi \right) -\frac{i}{\phi }\left(
\delta \partial _{\mu }\bar{\psi}\right) \beta ^{\mu }\gamma \psi \right] 
\notag \\
-\int_{\Omega }dx~e\phi ^{4}\left( i\nabla _{\mu }\bar{\psi}\beta ^{\mu
}\gamma +\bar{\psi}\gamma -i\bar{\psi}\gamma \beta ^{\mu }S_{\mu }\right)
\delta \psi  \notag
\end{gather}%
Choosing different specializations of the variations, one can easily obtain
the equations of motion and the energy-momentum and spin density tensor.

\subsection{Equations of Motion}

We choose to make functional variations only in the massless DKP field thus
we set $\delta \phi =\delta e^{\mu }{}_{b}=\delta \omega _{\mu ab}=0$ and
considering $\left[ \delta ,\partial _{\mu }\right] =0$, from ( \ref%
{action-variation}) we get%
\begin{gather*}
\delta S=\int_{\partial \Omega }d\sigma _{\mu }~e\phi ^{3}~i\left[ \frac{{}}{%
{}}\bar{\psi}\gamma \beta ^{\mu }\left( \delta \psi \right) -\left( \delta 
\bar{\psi}\right) \beta ^{\mu }\gamma \psi \right] + \\
+i\int_{\Omega }dx~e\phi ^{4}\left( \delta \bar{\psi}\right) \left[ i\beta
^{\mu }\nabla _{\mu }\psi +i\tilde{\tau}_{\mu }\beta ^{\mu }\gamma \psi
-\gamma \psi \frac{{}}{{}}\right] + \\
-\int_{\Omega }dx~e\phi ^{4}\left[ i\nabla _{\mu }\bar{\psi}\beta ^{\mu }+i%
\tilde{\tau}_{\mu }\bar{\psi}\gamma \beta ^{\mu }+\bar{\psi}\gamma \frac{{}}{%
{}}\right] \delta \psi
\end{gather*}

Following the action principle we get the generator of the variations of the
massless DKP\ field%
\begin{equation*}
G_{\delta \psi }=\int_{\partial \Omega }d\sigma _{\mu }~e\phi ^{3}~i\left[ 
\frac{{}}{{}}\bar{\psi}\gamma \beta ^{\mu }\left( \delta \psi \right)
-\left( \delta \bar{\psi}\right) \beta ^{\mu }\gamma \psi \right]
\end{equation*}%
and its equations of motion in the Lyra's manifold are%
\begin{equation*}
\left\{ 
\begin{array}{c}
i\beta ^{\mu }\left( \nabla _{\mu }+\tilde{\tau}_{\mu }\gamma \right) \psi
-\gamma \psi =0 \\ 
i\nabla _{\mu }\bar{\psi}\beta ^{\mu }+i\tilde{\tau}_{\mu }\bar{\psi}\gamma
\beta ^{\mu }+\bar{\psi}\gamma =0%
\end{array}%
\right.
\end{equation*}

The spin 1 projectors $R^{\mu }$($=e^{\mu }{}_{a}R^{a}$ ) and $R^{\mu \nu }$(%
$=e^{\mu }{}_{a}e^{\nu }{}_{b}R^{ab}$) \cite{Umezawa, Lunardi1} are such
that $R^{\mu }\psi $ and $R^{\mu \nu }\psi $ transform respectively as a
vector and a second rank tensor under general coordinate transformation.
Thus, using the projectors we have 
\begin{equation*}
R^{\mu }\,\ \ \rightarrow \,\ \ i\nabla _{\nu }\left( R^{\mu \nu }\psi
\right) +i\tilde{\tau}_{\nu }\left( R^{\mu \nu }\gamma \psi \right) -R^{\mu
}\gamma \psi =0
\end{equation*}%
multiplying by $\left( 1-\gamma \right) $ we get 
\begin{equation*}
i\left( \nabla _{\nu }+\tilde{\tau}_{\nu }\right) \left( R^{\mu \nu }\gamma
\psi \right) =0
\end{equation*}%
and 
\begin{equation*}
R^{\mu \nu }\,\ \ \rightarrow \,\ \ i\nabla _{\rho }\left( R^{\mu \nu }\beta
^{\rho }\psi \right) +i\tilde{\tau}_{\rho }\left( R^{\mu \nu }\beta ^{\rho
}\gamma \psi \right) -R^{\mu \nu }\gamma \psi =0
\end{equation*}

\begin{equation*}
R^{\mu \nu }\gamma \psi =i\left( \nabla _{\rho }+\tilde{\tau}_{\rho }\gamma
\right) \left[ g^{\rho \nu }\left( R^{\mu }\psi \right) -g^{\rho \mu }\left(
R^{\nu }\psi \right) \right]
\end{equation*}%
from the above equations we get the equation of motion for the massless
vector field $R^{\mu }\psi $ 
\begin{equation*}
(\nabla _{\nu }+\tilde{\tau}_{\nu })(\nabla _{\rho }+\tilde{\tau}_{\rho
}\gamma )\left[ g^{\rho \nu }\left( R^{\mu }\psi \right) -g^{\rho \mu
}\left( R^{\nu }\psi \right) \right] =0\,,
\end{equation*}

We use a specific representation of the DKP algebra in which the singular $%
\gamma $ matrix is 
\begin{equation*}
\gamma =\mbox{diag}(0,0,0,0,1,1,1,1,1,1)~.
\end{equation*}%
Then in this representation the DKP\ field $\psi $ is now a 10-component
column vector 
\begin{equation*}
\psi =\left( 
\begin{array}{c}
\psi ^{0},\psi ^{1},\psi ^{2},\psi ^{3},\psi ^{23},\psi ^{31},\psi
^{12},\psi ^{10},\psi ^{20},\psi ^{30}%
\end{array}%
\right) ^{T}\;,
\end{equation*}%
where $\psi ^{a}$ ($a=0,1,2,3$) and $\psi ^{ab}$ behave, respectively, as a
4-vector and an antisymmetric tensor under \textit{Lorentz} transformations
on the Minkowski tangent space. And we also get

\begin{gather*}
\gamma \psi =\left( 
\begin{array}{c}
0,0,0,0,\psi ^{23},\psi ^{31},\psi ^{12},\psi ^{10},\psi ^{20},\psi ^{30}%
\end{array}%
\right) ^{T} \\
R^{\mu }\psi =\left( 
\begin{array}{c}
\psi ^{\mu },0,0,0,0,0,0,0,0,0%
\end{array}%
\right) ^{T} \\
R^{\mu \nu }\psi =\left( 
\begin{array}{c}
\psi ^{\mu \nu },0,0,0,0,0,0,0,0,0%
\end{array}%
\right) ^{T}
\end{gather*}%
due to $R^{\mu }\gamma =\gamma R^{\mu }$ and $R^{\mu \nu }\gamma =(1-\gamma
)R^{\mu \nu }$. Then, we get the following relations among $\psi $
components 
\begin{equation*}
i\,\psi _{\mu \nu }=\nabla _{\mu }\psi _{\nu }-\nabla _{\nu }\psi _{\mu }
\end{equation*}%
which leads to the equation of motion for the spin $1$ sector of the
massless DKP field in Lyra space-time 
\begin{equation*}
(\nabla _{\mu }+\tilde{\tau}_{\mu })\left( {\nabla }^{\mu }\psi ^{\nu }-{%
\nabla }^{\nu }\psi ^{\mu }\right) =0\,.
\end{equation*}

\subsection{Energy-momentum tensor and spin tensor density}

Now, we only vary the background manifold and we assume that $\delta \omega
_{\mu ab}$\ and $\delta e^{\mu }{}_{a}$\ are independent variations, 
\begin{gather*}
\delta S=\int_{\Omega }dxe\phi ^{4}\left[ i\left( \bar{\psi}\gamma \beta
^{a}\nabla _{\mu }\psi -\nabla _{\mu }\bar{\psi}\beta ^{a}\gamma \psi
\right) \delta e^{\mu }{}_{a}+\right. \\
\left. +\left( \frac{1}{e}\delta e\right) \mathcal{L}+i\left( \bar{\psi}%
\gamma \beta ^{\mu }S^{ab}\psi +\bar{\psi}S^{ab}\beta ^{\mu }\gamma \psi
\right) \frac{1}{2}\delta \omega _{\mu ab}\right] .
\end{gather*}

First, holding only the variations in the tetrad field, $\delta \omega _{\mu
ab}=0$, we found for the variation of the action 
\begin{equation*}
\delta S=\int_{\Omega }dx~e\phi ^{4}~\left[ \frac{{}}{{}}i\left( \bar{\psi}%
\gamma \beta ^{a}\nabla _{\mu }\psi -\nabla _{\mu }\bar{\psi}\beta
^{a}\gamma \psi \right) -e_{\mu }{}^{a}\mathcal{L}\right] \delta e^{\mu
}{}_{a}
\end{equation*}

Defining the energy-momentum density tensor as 
\begin{equation*}
T_{\mu }{}^{a}\equiv \frac{1}{\phi ^{4}e}\frac{\delta S}{\delta e^{\mu
}{}_{a}}=i\bar{\psi}\gamma \beta ^{a}\nabla _{\mu }\psi -i\nabla _{\mu }\bar{%
\psi}\beta ^{a}\gamma \psi -e_{\mu }{}^{a}\mathcal{L}
\end{equation*}%
which can be written in coordinates as 
\begin{equation*}
T_{\mu }{}^{\nu }\equiv e^{\nu }{}_{a}T_{\mu }{}^{a}=i\bar{\psi}\gamma \beta
^{\nu }\nabla _{\mu }\psi -i\nabla _{\mu }\bar{\psi}\beta ^{\nu }\gamma \psi
-\delta _{\mu }{}^{\nu }\mathcal{L}
\end{equation*}%
On the mass shell, 
\begin{equation*}
T_{\mu }{}^{\nu }=i\bar{\psi}\gamma \beta ^{\nu }\nabla _{\mu }\psi -i\nabla
_{\mu }\bar{\psi}\beta ^{\nu }\gamma \psi -\delta _{\mu }{}^{\nu }\bar{\psi}%
\gamma \psi
\end{equation*}

Now, making functional variations only in the components of the spin
connection, $\delta e^{\mu }{}_{a}=0$, we found for the action variation 
\begin{equation*}
\delta S=\int_{\Omega }dx~e\phi ^{4}~\frac{1}{2}\left( \delta \omega _{\mu
ab}\right) i\bar{\psi}\left( \gamma \beta ^{\mu }S^{ab}+S^{ab}\beta ^{\mu
}\gamma \right) \psi ,
\end{equation*}%
we define the spin tensor density as being 
\begin{equation*}
S^{\mu ab}\equiv \frac{2}{\phi ^{4}e}\frac{\delta S}{\delta \omega _{\mu ab}}%
=i\bar{\psi}\left( \gamma \beta ^{\mu }S^{ab}+S^{ab}\beta ^{\mu }\gamma
\right) \psi
\end{equation*}

The spin 1 component of DKP energy momentum tensor is%
\begin{eqnarray*}
T_{\mu }{}^{\nu } &=&\frac{i}{2}\,\psi ^{\ast \nu \alpha }\left( \nabla
_{\mu }\,\psi _{\alpha }-\nabla _{\alpha }\,\psi _{\mu }\right) + \\
&&-\frac{i}{2}\,\psi ^{\nu \beta }\left( \nabla _{\mu }\,\psi _{\beta
}^{\ast }-\nabla _{\beta }\,\psi _{\mu }^{\ast }\right) + \\
&&-\delta _{\mu }{}^{\nu }\left( \psi ^{\ast \alpha \beta }\psi _{\alpha
\beta }\right)
\end{eqnarray*}%
which coincides with the first order energy momentum tensor of the
electromagnetic field in the real case.

\section{Final Remarks}

The coupling between torsion and massless vectorial field was showed to be
related to scale transformations in Lyra background. Since this scale
transformations are governed by an arbitrary function $\phi $, it seems
plausible that the problem of breaking the local gauge invariance associated
with this coupling could be removed from the theory if we had chosen an
gauge transformations to be linked to scale invariance in Lyra manifold. A
deeper study of this line is under construction.

\begin{center}
\textbf{Acknowledgments}
\end{center}

This work is supported by FAPESP grants 01/12611-7 (RC), 01/12584-0 (CAMM)
and 02/00222-9 (BMP). BMP also thanks CNPq for partial support.


\end{document}